\documentstyle[12pt,fleqn]{article}

\begin{document}

\title{ Exotic heavy baryons at LHC?\thanks{Work supported by BMFT, GSI
Darmstadt}
      }
\author{T.S.Bir\'{o}\\ {\small \em University of Giessen} \\
			{\small \em Heinrich-Buff-Ring 16, D-6300 Giessen, Germany }
            \\ and \\
        J.Zim\'anyi \\ {\small \em Central Research Institute for Physics}
					\\ {\small \em H-1525 Budapest 114, P.O.Box 49, Hungary}
        }
%\vspace*{-2.0cm}
\maketitle
%\vspace{3.0cm}
\begin{abstract}
\par
We speculate about a heavy bottom-charm six-quark baryon.
A semiclassical and a gaussian estimate reveal that the octet-octet
bbb-ccc configuration can be energetically favored with respect to
the singlet-singlet one.
This result suggests that a confined bbb-ccc six-quark state may exist.
Such objects may be produced in appreciable amount in heavy ion
collisions at LHC energies.
\end{abstract}
\newpage
\par
It has been recently considered in a study of high-energy heavy-ion collisions
that multi-heavy baryons may be abundantly produced if the hadronization
mechanism is described in the framework of a quark combinatoric model,
which includes a phenomenlogical penalty factor for charm and bottom
flavors [1].
In the present letter we speculate about a confined bottom-charm six-quark
system (multi-heavy dibaryon), which may be in an octet-octet color state
of the bbb core and ccc shell quarks, respectively.
Such 'higher order' \, heavy baryons could not necessarily be observed
up to now, since the heavy quarks decay weakly into light quarks destroying
the (possible) bound state of six quarks into two 'normal' \, baryons.
The only way to produce them seems to be a relativistic heavy-ion
collision in the LHC energy range.
\par
It is challenging also on its own right to investigate non-minimal color
multiplet arrangements of several quark systems in order to gain a new
insight into the nature of strong color forces in the non-relativistic
(heavy quark) regime and learn so about nonperturbative and slightly
collective QCD systems as well.
\par
In this letter we estimate the energy of the six-quark bbb-ccc system
treating the bbb core as one very heavy particle, which can either be in a
color octet or in a color singlet state being the total six-quark system
a color singlet. This way we compare the relative strength of the
singlet and octet confining force as well - a problem also addressed in
lattice QCD potential calculations [2] and recently in heavy meson physics
[3].
\par
The basis of our estimate is a simple non-relativistic potential model [4]
of heavy quark systems, like the $J/\psi$ meson. The Hamiltonian to be
used has the general form
\begin{equation}
H = \sum_i m_i - \sum_i \frac{1}{2m_i} \Delta_i - \sum_{i<j}
\vec{q}_i \cdot \vec{q}_j \left( \tilde{\sigma} r_{ij} -
\frac{\tilde{\alpha}}{r_{ij}} \right),
\end{equation}
where $m_i$ is the quark rest mass (we use m = 1.32 GeV for the charm quark),
$\Delta_i$ is the Laplace operator acting on the coordinate $\vec{r}_i$ of
the i-th quark, $r_{ij}$ is the relative distance of the quarks $i$ and $j$
and finally $\vec{q}_i \cdot \vec{q}_j$ is a symbolic notation for an
Ising-type SU(3) color-color interaction to be discussed in more detail later.
$\tilde{\sigma}$ and $\tilde{\alpha}$ are phenomenological potential
parameters, which we can connect to the string tension $\sigma$ and
color coupling strength $\alpha$ used in heavy-meson physics by applying
the above Hamiltonian (1) to the $J/\psi$ system.
\par
Before presenting some estimates we would like to emphasize that the key
physical reason to expect
the existence of heavy multibaryons is that
in a heavy quark system the quarks are much nearer to each other than
in light quark systems and therefore they feel much stronger attractive
color force than the light quarks.
Furthermore, the color charges belonging
to higher multiplets amplify the confining (attractive) forces between the
quarks. Having possibly heavy valence quarks in mind the increase in the
kinetic energy because of more degrees of freedom in the relative motion
can be minor in comparison to this effect favoring so bound multibaryon
states, such as a three-charm cloud around a compact bbb core would be.
\par
Let us enlighten this reasoning by investigating some special quark
configurations in this model: the $J/\psi$ ( a charm quark and an
antiquark belonging to a $3$ and an $\bar{3}$ SU(3) color representation
and being alltogether a singlet),
the $\Omega_c$ (three charm quarks each belonging to a $3$ and forming
a singlet alltogether) and the hypothetical bbb-ccc dibaryon.
In the latter case we consider only four color charges: the compact bbb
core and each of the relatively light charm quarks.
The total system being color neutral and three quarks giving together either
a singlet, an octet or a decuplet representation we deal with the
following possibilities:
\begin{enumerate}
\item   The bbb core is a singlet. In this case the ccc system is decoupled and
                must be a singlet so that a c quark ($3$) is attracted by a cc
                diquark $\bar{3}$.
\item   The bbb core is an octet. Now the ccc system is also an octet and
                the cc diquark can either be an antitriplet or a sextet, since
both
                $3 \times \bar{3}$ and $3 \times 6$ contains an octet.
\end{enumerate}
\par
Generally, considering the color symmetry inherent in the QCD, we may
distinguish between color configurations by grouping different quarks
into a given multiplet as long as the whole system is color neutral,
but we are not allowed to assign different energy to different color
projections (i.e. to a quark state which is actually 'red').
Therefore we determine the color Ising factors $\vec{q}_i \cdot \vec{q}_j$
for each case formally from the color charge square Casimir operator of
the total charge and the charge of the subsystems under consideration.
Doing so we use the general formula
\begin{equation}
 \sum T^a T^a = Q^2 \cdot 1,
\end{equation}
where  $1$ is the unit matrix, the $T^a$-s are the generators of the SU(3)
algebra in the corresponding representation and the eigenvalue of the
Casimir operator in a $(J,j)$ SU(3) multiplet is given by [5]
\begin{equation}
 Q^2 = \frac{1}{9} \left( J^2 + j^2 - J j + 3 J \right).
\end{equation}
For a singlet, triplet or antitriplet, sextet and octet we get
0, 4/9, 10/9 and 1 for $Q^2$, respectively. For the $J/\psi$ meson we obtain
this way the color Ising factor
\begin{equation}
 \vec{q}_1 \cdot \vec{q}_2 = \frac{1}{2} \left( \left( \vec{q}_1 + \vec{q}_2
 \right)^2 - q_1^2 - q_2^2 \right) = - \frac{4}{9},
\end{equation}
leading to the energy formula
\begin{equation}
 E^{J/\psi} = 2m + \frac{P^2_0}{m} + \frac{4}{9} f(r_{12}),
\end{equation}
with
\begin{equation}
 f(r) = \tilde{\sigma} r - \frac{\tilde{\alpha}}{r}.
\end{equation}
The kinetic energy of the relative motion can be estimated semiclassically
using
\begin{equation}
 \frac{P_0^2}{m} = \frac{K}{mr^2}.
\end{equation}
{}From a fit to $J/\psi$ [6], which uses $\sigma = 0.192$ GeV, $\alpha = 0.47$
and $m = 1.32$ GeV we obtain $K = 1.39$ leading to the experimental
$J/\psi$ mass $E = 3.096$ GeV. This value we use for other heavy quark systems
in our first estimate as well.
\par
For the bbb-ccc system we get three different contributions to the kinetic
energy (ccc's relative motion to the bbb core, a c quark's motion relative to
the cc diquark and finally a relative motion inside the diquark) with the
respective reduced masses. Assuming a symmetric configuration,
$r_{12}=r_{23}=r_{31}=r$ we obtain
\begin{equation}
 E_{bbb-ccc}^{kin} = \frac{K}{6 m r_+^2} +
        \frac{K}{\frac{4}{3} m r_-^2} + \frac{K}{m r_0^2}.
\end{equation}
belonging to the Jacobian coordinates [7] defined as
\begin{eqnarray}
 \vec{r}_+ &=& \frac{1}{3} \left( \vec{r}_1 + \vec{r}_2 + \vec{r}_3 \right)
\nonumber \\ \nonumber \\
 \vec{r}_- &=& \vec{r}_3 - \frac{1}{2} \left( \vec{r}_1 + \vec{r}_2 \right)
\nonumber \\ \nonumber \\
 \vec{r}_0 &=& \vec{r}_1 - \vec{r}_2.
\label{Jacobi}
\end{eqnarray}
We consider
the color Ising factors relevant for the potential energy in this system
first with a color permutation assumption leading to
\begin{equation}
 \vec{Q}\cdot\vec{q}_i = - \frac{1}{3} Q^2
\end{equation}
and
\begin{equation}
 \vec{q}_i \cdot \vec{q}_j = \frac{1}{6} \left( Q^2 - \frac{4}{3} \right)
\end{equation}
for each corresponding bbb-c or c-c quark pair. For the singlet core $Q^2=0$,
for the octet one $Q^2=1$ must be taken, so we arrive at
\begin{equation}
 V^1 = \frac{2}{9} \left( f(r_{12}) + f(r_{23}) + f(r_{31}) \right)
\end{equation}
for the singlet and
\begin{equation}
 V^8 = \frac{1}{18} \left( f(r_{12}) + f(r_{23}) + f(r_{31}) \right)
        + \frac{1}{3} \left( f(r_1) + f(r_2) + f(r_3) \right)
\end{equation}
for the octet core.
\par
A symmetric triangle configuration of the c quarks with a singlet bbb core
$(r_+ = \infty, r_0=r, r_-=r\sqrt{3}/2)$ leads to a minimal energy of
$E^1=4.71$ GeV at $r=2.15$ GeV$^{-1}$, while an equal tetraeder with an
octet bbb $(r_+=r\sqrt{2/3}, r_-=r\sqrt{3}/2, r_0=r)$ leads to
$E^8=4.78$ GeV at $r=1.6$ Gev$^{-1}$, which is just slightly above the
singlet energy.
\par
This result let us be optimistic about to make a more advanced estimate.
First, we distinguish now between the antitriplet ($(\vec{q}_1+\vec{q}_2)^2 =
q^2 = 4/9$) and the sextet ($q^2 = 10/9$) cc diquark states. With this
distinction and assuming that the triplet c quark is located at $\vec{r}_3$
while the diquark consists of the quarks located at $\vec{r}_1$ and $\vec{r}_2$
respectively we arrive at the following classification of the color
Ising interaction.
\par
The bbb core - c quark interaction energy is nonzero only if the core is
in an octet state. In this case either
\begin{equation}
 V_0^{8,6} = \frac{5}{12} f(r_1) + \frac{5}{12} f(r_2) + \frac{1}{6} f(r_3),
\end{equation}
if the diquark is in a sextet or
\begin{equation}
 V_0^{8,\bar{3}} = \frac{1}{4} f(r_1) + \frac{1}{4} f(r_2) + \frac{1}{2}
f(r_3),
\end{equation}
if the diquark is in an antitriplet state.
The c - cc interaction energy is
\begin{equation}
 V_3^{8,6} = \frac{5}{36} \left( f(r_{23}) + f(r_{31}) \right)
\end{equation}
and
\begin{equation}
 V_3^{8,\bar{3}} = - \frac{1}{36} \left( f(r_{23}) + f(r_{31}) \right)
\end{equation}
for an octet core and
\begin{equation}
 V_3^{1,\bar{3}} = \frac{2}{9} \left( f(r_{23}) + f(r_{31}) \right)
\end{equation}
for a singlet core, respectively.
Finally the intra-diquark interaction is either
\begin{equation}
 V_{12}^{\bar{3}} = \frac{2}{9} f(r_{12})
\end{equation}
for an antitriplet or
\begin{equation}
 V_{12}^{6} = - \frac{1}{9} f(r_{12})
\end{equation}
for a sextet diquark state.
For the sake of comparison we give here the interaction energy inside the
$J/\psi$ meson, where the $c - \bar{c}$ system is in a singlet state, as well
\begin{equation}
 V_{12}^{1} = \frac{4}{9} f(r_{12}).
\end{equation}
Besides noting the color hyperfine splitting of the interaction energies an
interesting frustration phenomenon known from spin glasses can be
observed: in the four color charge system with an octet bbb core cannot
be all pairwise interaction attractive (and hence confining), although
the whole system is confined.
\par
The second improvement on the semicalssical estimate is to use a gaussian
wave function ansatz,
\begin{equation}
 \Psi \propto \exp \left( - \left( 9r_+^2/R_+^2 + r_-^2/R_-^2 + r_0^2/4R_0^2
  \right) / 4 \right),
\end{equation}
with the Jacobian coordinates defined in eq.(\ref{Jacobi}).
A gaussian ansatz, being with some calculational tricks always factorizable
in the very coordinate needed to evaluate the interaction energy, leads to
the generic result
\begin{equation}
 < \frac{1}{r} > = \frac{4}{\pi} \frac{1}{<r>},
\end{equation}
so we need to present the gaussian averages of  c-c and c-bbb quark distances
only. They are
\begin{eqnarray}
 <r_{12}> &=&  4 \sqrt{\frac{2}{\pi}} R_0
\nonumber \\ \nonumber \\
 <r_{23}> = <r_{31}> &=& \frac{2}{\sqrt{\pi}} \left( R_0^2 + R_-^2
\right)^{1/2}
\nonumber \\ \nonumber \\
 <r_3> &=& \frac{4}{3} \frac{1}{\sqrt{2\pi}} \left( R_+^2 + 4R_-^2
\right)^{1/2}
\nonumber \\ \nonumber \\
 <r_1> = <r_2> &=& \frac{4}{3} \frac{1}{\sqrt{2\pi}}
  \left( R_+^2 + R_-^2 + 9R_0^2 \right)^{1/2}.
\end{eqnarray}
Minimizing the energy numerically in terms of the ansatz parameters $R_+$,
$R_-$ and $R_0$ we obtain the following energies using the standard
parameters of ref. [6], $\sigma = 0.192$ GeV$^2$, $\alpha=0.47$:
\begin{tabbing}
dsfafgcgvjhghjj \= \qquad \qquad hhadhhdajjlka \kill \\
$J/\psi$:                               \>      \qquad 3.09 GeV \\
$\Omega_c$:                             \>      \qquad 4.54 GeV \\
$bbb-ccc (8,6)$:                \>      \qquad 4.57 GeV \\
$bbb-ccc (8,\bar{3})$:  \>      \qquad 4.74 GeV \\
\end{tabbing}
not counting the 15 GeV bbb rest mass in the two letter cases.
Although this estimate slightly disfavors the confined bbb-ccc system
we do think that this problem is worth of further investigation, because
there is some freedom in choosing the parameters  used in this estimate.
In fact varying the coupling strength $\alpha$ at fixed string tension
$\sigma = 0.192$ GeV$^2$ a critical range around $\alpha = 0.5$
can be identified, where the bbb-ccc system may or may not be
energetically favored (fig.1).
\par
On the other hand, if we speculate about an altered string constant in
many-quark systems, using for example $\sigma = 0.22$ GeV$^2$ and
$\alpha = 0.54$ we obtain
\begin{tabbing}
dsfafgcgvjhghjj \= \qquad \qquad hhadhhdajjlka \kill \\
$J/\psi$:                               \>      \qquad 3.097 GeV \\
$\Omega_c$:                             \>      \qquad 4.544 GeV \\
$bbb-ccc (8,6)$:                \>      \qquad 4.515 GeV \\
$bbb-ccc (8,\bar{3})$:  \>      \qquad 4.716 GeV \\
\end{tabbing}
\par
Concluding this letter we investigated whether the heavy bbb-ccc
six-quark system can be energetically favored with respect to two
separate heavy baryons. Although we have not yet found a decisive
quantitative result, a semiclassical and a gaussian estimate show that
this possibility cannot be excluded on the basis of phenomenological
knowledge about heavy meson masses, like $J/\psi$.
Since such exotic heavy baryons are in principle producable in
relativistic heavy ion collisions at LHC energies and they may be
observed through a  measurement of like-charge muon
triplets [8], we do think that a concise future investigation
of this problem with more advanced calculations is desirable.
\vspace{1cm}
\par
{\bf Acknowledgement:}
The authors acknowledge the kind hospitality of GSI offered by W. N\"orenberg,
where part of this work has been done. This work was also supported by
OTKA (National Scientific Research Fund, Hungary) under Grant. No. 2973
and by BMFT (Bundesministerium f\"ur Forschung und Technologie, Germany).

\par
%\begin{large}
%{\bf References}
%\end{large}

\vspace{1.0cm}
\par
\begin{large}
{\bf Figure caption}
\end{large}
\begin{itemize}
\item[Fig.1.]   The phenomenlogical energy of a $c\bar{c}$ and a $ccc$
        heavy quark system in a singlet or in an octet state bounded to an
        octet $bbb$ core.
\end{itemize}

\end{document}